\documentclass{aa}  
\usepackage{graphicx}
\usepackage{txfonts}

\usepackage[T1]{fontenc}

\usepackage{graphicx}	% Including figure files
\usepackage{amsmath}	% Advanced maths commands
\usepackage{acronym}
\usepackage{comment}    % (remove for final version)
\usepackage{hyperref}

\def\half{\frac{1}{2}}

\def\Msun{{\rm M_{\odot}}}
\def\GMc2{{\rm G M_{\odot} c^{-2}}}
\def\Mpc{{\rm Mpc}}

\def\M{\mathcal{M}}
\def\B{\mathcal{B}}

\def\Mo{{\rm M_{\odot}}}
\def\kt2{\kappa^\text{T}_2}
\def\Mmax{M_\text{max}^\text{TOV}}
\def\Rmax{R_\text{max}^\text{TOV}}
\def\R14{R_\text{1.4}^\text{TOV}}
\def\Rmax{R_\text{max}^\text{TOV}}

\def\lambdat{\tilde\Lambda}
\def\tLam{\lambdat}
\def\rhosat{\rho_{\rm sat}}
\def\pr{{\rm p}}
\def\data{\boldsymbol{d}}
\def\params{\boldsymbol{\theta}}
\def\tref{{t_{\rm mrg}}}
\def\pref{{\phi_{\rm mrg}}}
\def\B{\mathcal{B}}
\def\mdisk{m_{\rm disk}}
\def\nrmap{\boldsymbol{F}_{\rm NR}}
\def\spin{\boldsymbol{\chi}}
\def\dyn{{\rm d}}
\def\wind{{\rm w}}
\def\mej{m_{\rm ej}}

\def\mej{M_\mathrm{ej}}

\def\eg{{\it e.g.}}
\def\ie{{\it i.e.}}
\def\cf{{\it cf.}}

\usepackage{pifont} % \cmark \xmark

\newcommand{\be}{\begin{equation}}
\newcommand{\ee}{\end{equation}}
\newcommand{\bea}{\begin{eqnarray}}
\newcommand{\eea}{\end{eqnarray}}
\newcommand{\bel}{\begin{align}}
\newcommand{\eel}{\end{align}}

\newcommand{\bajes}[1]{{\tt bajes{#1}}}
\newcommand{\TEOB}{{\tt TEOBResumS}}

\usepackage{color}
\definecolor{cyan}{rgb}{0,0.9,0.9}
\definecolor{orange}{rgb}{0.9,0.5,0}
\definecolor{magenta}{rgb}{1,0,1}
\definecolor{purple}{rgb}{0.8,0.4,0.8}
\definecolor{gray}{rgb}{0.8242,0.8242,0.8242}
\acrodef{BNS}[BNS]{binary neutron star}
\acrodef{NR}[NR]{Numerical relativity}
\acrodef{GR}[GR]{general relativity}
\acrodef{EoS}[EOS]{equation of state}
\acrodef{GW}[GW]{gravitational wave}
\acrodef{EM}[EM]{electromagnetic}
\acrodef{MHD}[MHD]{magnetohydrodynamics}
\acrodef{NS}[NS]{neutron star}
\acrodef{BH}[BH]{black hole}
\acrodef{LK}[LK]{leakage}
\acrodef{GRB}[SGRB]{short-gamma-ray burst}
\acrodef{GRLES}[GRLES]{general relativistic Large-Eddy-Simulations}

\begin{document} 

\title{Bayesian inference of multimessenger astrophysical data:\\
  Joint and coherent inference of gravitational waves and kilonovae}

\titlerunning{\bajes{-MMA}}
\authorrunning{Breschi et al.}

\author{
  Matteo Breschi \inst{1,2,3}
  \and Rossella Gamba \inst{1,4,5}
  \and Gregorio Carullo \inst{1,6}
  \and Daniel Godzieba \inst{4}
  \and \\
  Sebastiano Bernuzzi \inst{1}
  \and Albino Perego \inst{7,8}
  \and David Radice \inst{4,9,10} \thanks{Alfred P.~Sloan Fellow}
}
\institute{
  % List of institutions
  Theoretisch-Physikalisches Institut,
  Friedrich-Schiller-Universit{\"a}t Jena, 
  Fr{\"o}belstieg 1, 07743 Jena, Germany 
  \and
  Scuola Internazionale Superiore di Studi Avanzati (SISSA), 
  via Bonomea 265, 34136 Trieste, Italy 
  \and
  Istituto Nazionale di Fisica Nucleare (INFN), Sezione di Trieste, 
  via Valerio 2, 34127 Trieste, Italy
  \and
  Institute for Gravitation \& the Cosmos, The Pennsylvania State University, University Park PA 16802, USA
  \and
  Department of Physics, University of California, Berkeley, CA 94720, USA
  \and
  Niels Bohr International Academy, Niels Bohr Institute, Blegdamsvej 17, 2100 Copenhagen, Denmark
  \and
  Dipartimento di Fisica, Università di Trento, Via Sommarive 14, 38123 Trento, Italy
  \and
  INFN-TIFPA, Trento Institute for Fundamental Physics and Applications, via Sommarive 14, I-38123 Trento, Italy
  \and
  Department of Physics, The Pennsylvania State University, University Park PA 16802, USA 
  \and
  Department of Astronomy \& Astrophysics, The Pennsylvania State University, University Park PA 16802, USA
}

% These dates will be filled out by the publisher
%\date{Accepted XXX. Received YYY; in original form ZZZ}
\date{\today}

\abstract
% context heading (optional)
    { Multimessenger observations of binary neutron star mergers can
      provide information on the neutron star's equation of state (EOS)
      above nuclear saturation density by directly constraining the
      mass-radius diagram.
    }
    % aims heading (mandatory)
    { We present a Bayesian framework for joint and coherent analyses of 
      multimessenger binary neutron star signals. As a first
      application, we analyze the gravitational-wave GW170817
      and the kilonova AT2017gfo data. These results are then combined
      with the most recent X-ray pulsars analyses of  
      PSR J0030+0451 and PSR J0740+6620 to obtain new EOS constraints.
    }
    % methods heading (mandatory)
    { We extend the \bajes{} infrastructure with a joint likelihood for multiple
      datasets, support for various semi-analytical kilonova models and
      numerical-relativity (NR) informed relations for the mass ejecta, as well
      as a technique to include and marginalize over modeling
      uncertainties. The analysis of GW170817 uses the \TEOB{} 
      effective-one-body waveform template to model the gravitational-wave
      signal. The analysis of AT2017gfo uses a baseline multi-component
      spherically symmetric model for the kilonova (kN) light curves.
      Various constraints on the mass-radius diagram and neutron star
      properties are then obtained by resampling over a set of ten 
      million parametrized EOS which is built under minimal assumptions (general relativity and causality). 
    }
    % results heading (mandatory)
    { We find that a joint and coherent approach improves the
      inference of the extrinsic parameters (distance)
      and, among the instrinc parameters, the mass ratio.
      The inclusion of NR informed relations
      strongly improves over the case of using an agnostic prior on
      the intrinsic parameters. 
      Comparing Bayes factors, we find that the two observations are
      better explained by the common source hypothesis only by assuming
      NR-informed relations. These relations break some of the
      degeneracies in the employed kN models.
      The EOS inference folding-in PSR J0952-0607 minimum-maximum mass,
      PSR J0030+0451 and PSR J0740+6620 data constrains, among other
      quantities, the 
      neutron star radius to $\R14={12.30}^{+0.81}_{-0.56}~{\rm km}$
      ($\R14={13.20}^{+0.91}_{-0.90}~{\rm km}$)
      and the maximum mass to $\Mmax={2.28}^{+0.25}_{-0.17}~{\Mo}$
      ($\Mmax={2.32}^{+0.30}_{-0.19}~{\Mo}$) where the 
      ST+PDT
      (PDT-U) analysis of Vinciguerra {\it et al} (2023) for PSR J0030+0451 is employed.
      Hence, the systematics on PSR J0030+0451 data reduction
      currently dominate the mass-radius diagram constraints.
    }
    % conclusions heading (optional), leave it empty if necessary 
    { We conclude that \bajes{} delivers robust analyses in-line with
      other stat-of-art results in the literature. Strong EOS constraints are
      provided by pulsars observations, albeit with large systematics
      in some cases. Current gravitational-wave constraints are
      compatible with pulsars constraints and can further improve the latter. 
    }
% Select between one and six entries from the list of approved keywords.
% Don't make up new ones.

\maketitle
\nolinenumbers

\section{Introduction}

The observation of the gravitational wave (GW) signal GW170817 from a binary
neutron star merger and its electromagnetic counterparts from the
merger aftermath opened new prospects to constrain the nature of
neutron star (NS) matter~\citep{VIRGO:2014yos, LIGOScientific:2014pky, TheLIGOScientific:2017qsa,Monitor:2017mdv,Abbott:2018wiz}.
GWs from the late-inspiral-to-merger frequencies carry
the imprint of short-range tidal interactions between the two
NSs~\citep{Damour:1986a,Damour:2009wj}. The measurement of tidal
polarizabilty parameters, in
particular the reduced tidal parameter $\tLam$ appearing in the
leading-order term of the GW phase~\citep{Damour:2012yf,Favata:2013rwa},  
constrains the nuclear equation of state (EOS) \citep{Abbott:2018exr,De:2018uhw,LIGOScientific:2019eut}. 
For GW170817, various analyses indicate posterior distributions
peaking in the interval 
$100\lesssim\tLam\lesssim800$, that can be mapped on a NS
radius~\footnote{%
As common in the literature, we call NS radius the radius $\R14$ of nonrotating, cold
beta-equilibrated NS of fiducial gravitational mass $1.4\Mo$. These general-relativistic
equilibrium configurations are calculated by solving the
Tolmann-Oppenheimer-Volkhoff equations. 
} constraint of $\R14=12.5^{+1.1}_{-1.8}$~km at 90\% credibility~\citep{Gamba:2020wgg}. 

Electromagnetic counterparts can complement such constraints by
delivering information on the merger remnant.
For example, the high-energy emission from a jet-like source of
GRB170817~\citep{Monitor:2017mdv,Savchenko:2017ffs} is associated to
the presence of a remnant black hole and can thus constrain the 
maximum NS mass~\citep{Margalit:2017dij,Shibata:2017xdx} (see also
\cite{Margalit:2022rde} for a revision of the first calculation.)
The intepretation of the kilonova (kN) 
AT2017gfo~\citep{Coulter:2017wya,Chornock:2017sdf,Nicholl:2017ahq,Cowperthwaite:2017dyu,Pian:2017gtc,Smartt:2017fuw,Tanvir:2017pws,Tanaka:2017qxj,Valenti:2017ngx}
requires different mass ejecta components
\eg~\citep{Abbott:2017wuw,Villar:2017wcc,Perego:2017wtu,Breschi:2021tbm},
thus excluding a prompt black hole formation \citep{Margalit:2017dij,Bauswein:2017vtn}.
Further, the necessity of including a massive disc wind to interpret
the light curves implies a lower limit (as opposite to the GW upper
limit) on the reduced tidal parameter~\citep{Radice:2017lry,Radice:2018pdn}.

The rigorous analysis of multi-messenger
astrophysics (MMA) data requires a Bayesian approach.
Bayesian inference of GW170817 and related counterparts with application to the NS EOS has been 
performed by various authors,
\eg~\citep{Radice:2018ozg,Coughlin:2018fis,Capano:2019eae,Coughlin:2019kqf,Jiang:2019rcw,Essick:2020flb,Dietrich:2020efo,Al-Mamun:2020vzu,Nicholl:2021abc,Breschi:2021tbm,Raaijmakers:2021uju,Ayriyan:2021prr,Huth:2021bsp,Pang:2022rzc,Brandes:2022nxa,Zhu:2022ibs,Fan:2023spm}. 
These analyses provide bounds or posterior distributions on fiducial NS
masses, radii, quadrupolar tidal polarizability parameters, as well as on
pressure (or energy density), sound speed at fiducial points or even
nuclear parameters (given a EOS paramtrization).
Some results about the NS radius are collected in Fig.~12 of
\cite{Breschi:2021tbm}, indicating a substantial agreement among
various analyses, with $\R14\simeq12$~km and $90\%$ credible intervals at the
kilometer level, depending on the specific assumptions. 
Note that, in several cases, these analyses incorporate assumption on
the EOS \citep{Jiang:2019rcw,Greif:2020pju,Huang:2023grj,Fan:2023spm} 
including available experimental nuclear data
\eg~\citep{Danielewicz:2002pu,Hebeler:2013nza,LeFevre:2015paj,Russotto:2016ucm}. 

Despite the potential of MMA observations of NS mergers, data from
GW170817 and counterparts alone do not significantly constrain the
nuclear physics of dense NS matter yet, \eg~\citep{Al-Mamun:2020vzu,Greif:2020pju}. 
The lower bound to the maximum NS mass $\Mmax$ (``minimum-maximum mass''), provided
by pulsars observations \citep{Demorest:2010bx,Miller:2019cac,Romani:2022jhd,Godzieba:2020tjn},
gives the strongest constraint on nuclear matter. Those data rule out a large number of EOS models, 
including some containing hyperons or deconfined quark matter \citep{Hebeler:2013nza} (though the latter
are still viable, see \eg~\cite{Annala:2019puf}).
The most massive NS identified so far is PSR J0952–0607, a millisecond
pulsar in a binary system, with
$M_{\rm J0740+6620}=2.35\pm0.17\Mo$~\citep{Romani:2022jhd}.
Recent X-ray observations of isolated pulsars 
have been performed by NICER and
XMM-Newton~\citep{Miller:2019cac,Riley:2019yda,Miller:2021qha,Riley:2021pdl}.  
These observations targeted
PSR J0030+0451~\citep{Miller:2019cac,Riley:2021pdl,Raaijmakers:2019qny} 
and PSR J0740+6620~\citep{Miller:2021qha}.
The former NS has a best radius and mass estimates of  
$R_{\rm J0030+0451}=12.71^{+1.14}_{-1.19}$~km and 
$M_{\rm  J0030+0451}=1.24^{+0.15}_{-0.16}\Mo$ (68\% credibility)~\citep{Miller:2019cac}
(see also~\citep{Riley:2019yda}).
The latter NS has the second-heaviest reliably determined mass to-date
$M_{\rm  J0740+6620}=2.08\pm0.07\Mo$
with a radius of $R_{\rm J0740+6620}=13.7^{+2.6}_{-1.5}$~km  (68\% credibility)
\citep{Miller:2021qha}. 
Overall, these data constrain the mass and radius 
of the NS at the ${\lesssim}5\%$ level.
Several MMA analyses have combined binary neutron star merger 
with pulsar data, \ie~used data from multiple sources, to improve
astrophysical EOS
constraints~\citep{Jiang:2019rcw,Essick:2020flb,Dietrich:2020efo,Al-Mamun:2020vzu,Nicholl:2021abc,Breschi:2021tbm,Raaijmakers:2021uju,Ayriyan:2021prr,Brandes:2022nxa,Fan:2023spm}.
Recently, \cite{Vinciguerra:2023qxq} re-analysed PSR J0030+0451 with an improved
pipeline, finding updated measurements of both the mass and radius of the pulsar
depending on the employed hot-spots model: $(M,R)_{\rm J0030+0451} =(1.4^{+
  0.13}_{- 0.12}~\Msun , 11.71^{+0.88}_{-0.83}~{\rm km})$ (ST+PDT) 
or  $(M,R)_{\rm J0030+0451} =(1.7^{+0.18}_{- 0.19}~\Msun ,14.44^{+ 0.88}_{-
  1.05}~{\rm km})$ (PDT-U).

Methodologically, MMA analyses share several common features.
We comment on three of such common elements, that are later incorporated in our work.
(i) The use of phenomenological relations from numerical-relativity (NR) simulations, 
incorporating remnant constraints into GW analysis. Accurate simulation
results are, for example, available to build equal-mass prompt collapse models, \eg 
\citep{Hotokezaka:2011dh,Bauswein:2013jpa,Agathos:2019sah,Kashyap:2021wzs,Perego:2021mkd}.
Simple models describing kinematic quantities of dynamical ejecta
\citep{Dietrich:2016fpt,Radice:2018pdn,Kruger:2020gig,Nedora:2020qtd} and remnant disc masses 
\citep{Radice:2018pdn,Kruger:2020gig,Nedora:2020qtd} in terms of the
binary properties are also available. However, such relationships are subject to
significant systematics, depending on the physics input of the
simulations~\citep{Nedora:2020qtd}. 
(ii) The assumption of a EOS catalog in order to map the 
posteriors of the inferred parameters into other set of parameters. The typical sample size of this catalog ranges from a few tens to a few thousands EOS curves. 
These EOS sets may be constructed from model-agnostic piecewise polytropic
representations
\eg~\citep{Jiang:2019rcw,Raaijmakers:2021uju} or assuming
nuclear theories, such as like chiral effective theory and perturbative quantum
chromodynamics in their regime of validity \eg~\citep{Essick:2020flb,Fan:2023spm}.
This implies different prior assumptions between different analyses,
and makes a direct comparison of the results problematic.
However, it should be noted that the vast majority of the employed EOS sets include 
constraints coming from massive pulsars~\citep{Antoniadis:2013pzd,Cromartie:2019kug}, 
sharing part of their prior information (additional details are provided below).
(iii) MMA analyses are often performed independently for each dataset,
combining in postprocessing the posterior distributions for the relevant
parameters. While this approach is justified for data coming from independent sources, 
the analysis of different data from a single source may benefit from joint coherent analyses,
especially in the case of large correlations between parameters describing the different dataset and in the presence of modelling systematics.
The \textit{single-source--multiple-data} scenario can be rigorously handled 
within the Bayesian framework by joinining the single messenger
likelihoods and performing a combined sampling of the full posterior probability distribution
\eg~\citep{Biscoveanu:2019bpy,Pang:2022rzc}; we refer to
these analyses as \textit{joint and coherent}.

In this work, we present a new framework for joint-and-coherent MMA
Bayesian analyses. We apply our framework to the case of GW170817 and
AT2017gfo and provide updated constraints on the NS EOS.
The structure and summary of the paper is as follows.
Sec.~\ref{sec:method} describes the methods employed in our
analysis, and presents the extension of the \bajes{}
pipeline~\citep{Breschi:2021tbm} to MMA data.
Sec.~\ref{sec:res} describes the results of applying \bajes{-MMA} on GW170817 
and AT2017gfo data. We employ a state-of-art effective-one-body (EOB)
template and a spherically symmetric multi-component semi-analytical
kN model. We compare single messenger analyses to joint \& coherent
analyses used either an agnostic prior on intrinsic parameters or a 
NR-informed prior on intrinsic parameters.
Sec.~\ref{sec:eos} discusses EOS constraints from our new
analyses. We use a set of ${\sim}10$~million parameterized EOS built under
minimal assumptions, namely assuming general relativity, causality and
a minimum-maximum mass of $2.09\Mo$~\citep{Romani:2022jhd}. We combine GW170817+At2017gfo
data with independent pulsar data and include, for the first time, the recent
re-analysis of~\cite{Vinciguerra:2023qxq}.
Conclusions and an Appendix on new fitting formulas close the paper.

\section{Methods}
\label{sec:method}

Our analyses are based on Bayesian probability, which delivers
information on the source parameters in terms of their 
posterior probability distributions, an accurate
characterization of the correlations among parameters and the
possibility of ranking different hypotheses to explain the 
data~\citep[\eg][]{Jeffreys:1939xee}.  
Given the observed data $\data$
and a set of parameters $\params$
that characterize a model for the data (hypothesis $H$), 
the information on the parameters is encoded in the posterior
distribution $\pr(\params|\data,H)$.
Using Bayes theorem, $\pr(\params|\data,H)$ can be computed
as the product of the likelihood function $\pr(\data|\params,H)$ and
the prior distribution  of the model parameters $\pr(\params|H)$.
The evidence $\pr(\data|H)$ is instead employed 
in the context of model selection in order to discriminate 
different models.
Given two different hypotheses, say $H_{\rm A}$ and $H_{\rm B}$,
the Bayes' factor (BF)
\begin{equation}
  \label{eq:bf}
  \B^{\rm A}_{\rm B} = \frac{\pr(\data|H_{\rm A})}{\pr(\data|H_{\rm B})}\,,
\end{equation}
encodes the support of the data in favoring hypothesis A against 
hypothesis B
(within the assumption of uniform prior on $H_{{\rm A},{\rm B}}$)
\footnote{In order to lighten the notation,
the explicit dependency of a statistical quantity $\pr(x|H)$
on the corresponding underlying hypotheses $H$ is 
made implicit when not necessary,
i.e. $\pr(x|H)\mapsto \pr(x)$.}.

For multimessenger astrophysics data, the high-dimensional parameters' posterior distribution can have
non-trivial correlations and multimodalities. 
Numerical stochastic methods are the essential tools in order to
perform parameter estimation (PE). 
We employ the {\bajes{}} pipeline~\citep[][]{Breschi:2021tbm}
together with the the nested sampling
algorithm~\citep[\eg][]{Skilling:2006,Feroz:2008xx}, implemented in the {\scshape dynesty} nested sampler~\citep[][]{Speagle:2020}.
All the PE runs presented here are performed with 5000 live points and
using {\bajes{}} parallel capabilities. In the following, we describe
the data, the likelihood functions, the models and the prior utilized
for the analyses.

\subsection{Gravitational waves inference}
\label{sec:method:gw}

The time-series recorded by the ground-based interferometers
LIGO and Virgo can be modeled as the sum of a noise contribution $n(t)$
and a GW transient $h(t)$, i.e. $d(t)=h(t)+n(t)$.
The signal observed by the interferometers
is computed from the GW polarizations $h_{+,\times}$ as 
\begin{equation}
  h(t)=F_+(\alpha,\delta,\psi) \, h_+(t)+
  F_\times(\alpha,\delta,\psi) \, h_\times(t)\,,
\end{equation}
where 
$F_{+,\times}$ are the antenna pattern functions
of the employed detector~\citep[see, \eg][]{Anderson:2000yy}
that are functions of the source location $\{\alpha,\delta\}$
and the polarization angle $\psi$.
We analyze the GW data segment $d(t)$
from the Gravitational-Wave Open Science Center (GWOSC)
centered around GPS time $1187008857$
with duration $T=128~{\rm s}$ and sampling rate
$4096$~Hz~\citep[][]{TheLIGOScientific:2017qsa,LIGOScientific:2018mvr}.

We analyze the GW signal $h(t)$ assuming a quasi-circular BNS merger and 
employing the effective-one-body (EOB) model
{\scshape TEOBResumS}~\citep[][]{Bernuzzi:2014owa,Nagar:2018zoe,Akcay:2018yyh,Nagar:2020pcj,Gamba:2020ljo}
with tidal and non-precessing spins interactions. We include the dominant 
quadrupolar $(2,2)$ mode of the radiation in the waveform construction, and use for efficiency the
post-adiabatic method \citep{Nagar:2018gnk} for the EOB dynamics and
the stationary-phase approximation for frequency domain
waveforms~\citep[][]{Gamba:2020ljo}.
We stress that this EOB model is faithful to NR within its error bands, and that systematic
errors due to modeling choices are sub-dominant with respect to statistical uncertainties 
for the GW analysis~\citep[][]{LIGOScientific:2018hze, Gamba:2020wgg}.

The GW template $h(t;\params_{\rm gw})$ is parameterized by 11 degrees of freedom, 
\begin{equation}
  \label{eq:gw-params}
  \params_{\rm gw} =\{m_1,m_2, \chi_1,\chi_2,\Lambda_1,\Lambda_2, 
  D_L^{gw}, \iota^{gw}, \psi, \tref^{gw}, \pref \}\,,
\end{equation}
where $m_{1,2}$ are the component masses, 
$\chi_{1,2}$ are the components of dimensionless spins 
aligned to the orbital angular momentum,
$\Lambda_{1,2}$ are the dimensionless quadrupolar tidal
polarizabilities, $D_L^{gw}$ is the luminosity distance,
$\iota^{gw}$ is the inclination angle between the line of sight and the 
total angular momentum of the system,
and $\tref^{gw}$ and $\pref$ are respectively the time and the phase at
merger. The GW supersript means that these parameters are the one associated with the GW signal, 
and the need for this distinction is made clear below.
For simplicity, the sky position is fixed to the location of the
optical counterpart~\citep[][]{GBM:2017lvd}, 
i.e. right ascension 13h 09m 48s 
and declination -23.3814 degrees.
Note that we assume $m_1\ge m_2$ and  
introduce the total binary mass $M=m_1+m_2$
and the mass ratio $q=m_1/m_2$.
The leading order tidal contribution
in the GW template is parameterized by the 
reduced tidal polarizability $\lambdat$ defined
as \citep{Damour:2009vw,Favata:2013rwa} 
\begin{equation}
  \label{eq:lambda-tilde}
  \lambdat = \frac{16}{13}\left[\frac{(m_1+12m_2)m_1^4}{M^5}\,\Lambda_1+(1\leftrightarrow 2)\right]\,.
\end{equation}

Data analyses of GW transients 
relies on the assumptions of stationarity and Gaussianity
of the noise in each detector, from which we can write a Gaussian likelihood function
in the Fourier domain as 
\begin{equation}
  \label{eq:gw-like}
  \begin{split}
    %	\log \pr(\data_{\rm gw}|\params_{\rm gw}) \propto -\half (d-h|d-h)\,,
    \log \pr(\data_{\rm gw}|\params_{\rm gw}) = 
    -\frac{2}{T} \sum_i &\frac{|\tilde d(f_i)-\tilde
      h(f_i)|^2}{S_n(f_i)}
    + \log\left[\frac{\pi T}{2} S_n(f_i)\right]\,,\\
  \end{split}
\end{equation}
where $\tilde h(f)$ is the Fourier transform of $h(t)$ (and analogously
for $\tilde d(f)$) while $S_n(f)$ is the power spectral density (PSD)
of the noise segment~\citep[][]{LIGOScientific:2018mvr}. 
The sum in Eq.~\eqref{eq:gw-like} is on the sampled frequencies and
evaluated over the frequency interval $[23~{\rm Hz}, 2~{\rm kHz}]$.
Note that \cite{Gamba:2020wgg} showed that GW analyses up to $2$~kHz are affected by
larger systematics than those at $1$~kHz on the tidal sector,
with the latter choice being more robust but also more
conservative. Since systematics effects are generically smaller than
other systematic effects discussed in this papers, we use here the
more commonly used $2$~kHz cut off.
Under the assumption that noise fluctuations
recorded in different detectors are not correlated,
the likelihood of the detector network is computed as the product of
the individual likelihoods.
We include spectral calibration envelopes 
with 10 logarithmically-spaced nodes
for each detector~\citep[see, \eg][]{Vitale:2011wu}.

The priors are the same as those discussed in \cite{Breschi:2021tbm}, 
with mass ratio bounded to $q\le 3$, 
isotropically-distributed spins $\spin_{1,2}$
constrained to $|\spin_{1,2}| \le 0.5$.

\subsection{Kilonovae inference}
\label{sec:method:kn}

We analyze the AT2017gfo AB magnitudes $\data_b(t)$ observed by various
telescopes in the photometric bands
$b=\{{\rm U}, {\rm B}, {\rm g}, {\rm V}, {\rm R}, 
{\rm I}, {\rm z}, {\rm J}, {\rm H}, {\rm K}, {\rm K}_{\rm s}\}$~\citep{Villar:2017wcc}.
The data provide a time coverage of ${\sim}20~{\rm days}$.
These data are provided with their associated standard deviations $\sigma_b$, and
corrected from reddening effects due to interstellar
extinction~\citep{Fitzpatrick:1998pb}. 

The kN model employed for our analyses is a multi-component
semi-analytical template for isotropic homologously-expanding ejecta
shells based on ~\citep{Grossman:2013lqa,Perego:2017wtu}.
Nuclear heating rates are described following \cite{Korobkin:2012uy,Barnes:2016umi}.
The model includes two ejecta components, each of which is
characterized by three parameters: 
the ejected mass $\mej$,
the root-mean-square velocity $v$,
and the gray opacity $\kappa$.
From a physical point of view, 
the less massive and fastest component may be associated to the
dynamical ejecta~\citep[\eg][]{Rosswog:2013,Radice:2016dwd}, 
while the slower ejecta component to baryonic winds radiated from the 
disk~\citep[\eg][]{Perego:2017xth,Radice:2018ghv}.
Thus, we label the first component as ``$\dyn$'' (i.e. dynamical ejecta)
and the second as ``$\wind$'' (i.e. baryonic wind).
Note, however, that in the inference we do not enforce specific information about the nature of these components,
and at the analysis level these just constitute labelling indices used to count the components. 
We prevent mode switching by ordering the components by decreasing velocity, with ``$\dyn$'' the fastest component.
The model is implemented and released in {\bajes{}}, which is designed
to host more complex semi-analytical models,~\citep[\eg][]{Perego:2017xth,Ricigliano:2023svx}.

Together with the ejecta parameters,
the kN light-curves $\ell_b(t)$
depend also on the extrinsic parameters of the source:
the luminosity distance $D_L^{kn}$,
the inclination angle $\iota^{kn}$ from the polar direction,
and the time $\tref^{kn}$ of coalescence.
Moreover, differently from \cite{Breschi:2021tbm}, we fix the heating
rate parameter
$\epsilon_0=2{\times}10^{18}~{\rm erg}~{\rm g}^{-1}~{\rm s}^{-1}$
according to \cite{Korobkin:2012uy}.
The AB magnitudes are parameterized by 9 degrees of freedom, 
\begin{equation}
  \label{eq:kn-params}
  \params_{\rm kn}=\{\mej^{\dyn}, v^{\dyn}, 
  \kappa^{\dyn}, \mej^{\wind}, v^{\wind}, \kappa^{\wind}, 
  D_L^{kn}, \iota^{kn}, t^{kn}_{\rm mrg} \}\,.
\end{equation}
Note that in our analysis, the assumption 
of isotropic kN removes the dependency on the inclination angle $\iota^{kn}$,
which in principle can be restored employing anisotropic ejecta
profiles~\citep{Breschi:2021tbm}.
However, in the following discussion on joint parameters, we will include this parameter for generality.

We assume that measurements performed at different times do not correlate
and introduce a Gaussian likelihood for each observed data
point~\citep[][]{Villar:2017wcc,Perego:2017wtu,Breschi:2021tbm}. 
Similarly to \cite{Villar:2017wcc}, we include an additional
correction to the data's standard deviation. These corrections are
inferred during the PE and are useful to mitigate 
systematic errors of the simple kN model used for the inference.
However, differently from  \cite{Villar:2017wcc}, we introduce a
correction $\Sigma_{b}$ for each photometric band
since the kN template can be diversely affected by systematic
errors at different electromagnetic wavelengths.
The kN likelihood is
\begin{equation}
  \label{eq:kn-like}
  \begin{split}
    \log \pr(\data_{\rm kn}|\params_{\rm kn}) = 
    -\half \sum_b  \sum_k& \,\frac{[d_b(t_k)-\ell_b(t_k)]^2}{\sigma_{b}^2(t_k)
      +\Sigma_{b}^2}\\
    &+ \log\left[2\pi \left(\sigma_{b}^2(t_k)+\Sigma_{b}^2\right)\right]\,,\\
  \end{split}
\end{equation}
where $\ell_b(t)$ represent the kN model
described above, $k$ runs over the
observed times.
Comparing to the GW likelihood, the kN likelihood has a different
normalization and it is smaller than one (hence the log-likelihood is
negative.) This is not problematic since the posteriors are normalised
and the evidence is always relative. 

The priors are taken uniform for $\mej^{(i)}\in [0,0.5]~\Mo$,
$v^{(i)}\in[0,0.333]~c$,
and $\kappa^{(i)}\in [0,50]~{\rm cm}^2~{\rm g}^{-1}$ for all components, with $i$ running on the components $d,w$. 
We do not impose any specific information about
the dynamical or wind nature of the component in the prior, \cf~\citep{Breschi:2021tbm}.
The  prior on the systematic deviations $\Sigma_b$
is taken as log-uniform constrained to $\log\Sigma_b\le 5 $~magnitudes.
This choice corresponds to the Jeffreys (uninformative) prior 
for standard deviation parameters of 
normal distributions~\citep[see][]{Jaynes:1968}.

\subsection{Joint \& coherent inference}
\label{sec:method:mm}

The likelihoods in Eq.~\eqref{eq:gw-like} and Eq.~\eqref{eq:kn-like}, 
together with the related prior assumptions,
provide a Bayesian framework for the inference of
the GW and kN parameters.
Within the assumption of different sources (``DS''),
the joined prior space is just the space product
of the GW and kN parameter spaces, 
\begin{equation}
  \label{eq:DS}
  \pr(\params_{\rm gw},\params_{\rm kn}|{\rm DS}) = 
  \pr(\params_{\rm gw})\,
  \pr(\params_{\rm kn})\,,
\end{equation}
There is no correlation between the GW and the kN parameters in
Eq.~\eqref{eq:DS}, 
\ie~$\params_{\rm gw} \cap \params_{\rm kn}=\emptyset$.
However, if the GW and kN transients are (assumed to be) originated 
from the same source, the spaces in
Eq.~\eqref{eq:gw-params} and Eq.~\eqref{eq:kn-params} share common
parameters. This implies a change of the prior and the sampling, as
discussed in the following. 

Observations from a single source (``SS'') are related by their 
extrinsic parameters, \ie~distance $D_L^{GW} = D_L^{KN}$, and similarly for the inclination $\iota$ and
merger time $\tref$ parameters.
To impose the above, we introduce the joint set of parameters:
\begin{equation}
  \label{eq:gw-params}
  \params_{\rm J} := \{m_1,m_2, \chi_1,\chi_2,\Lambda_1,\Lambda_2, \psi, \pref, \mej^{\dyn}, v^{\dyn}, 
  \kappa^{\dyn}, \mej^{\wind}, v^{\wind}, \kappa^{\wind}, D_L^{J}, \iota^{J}, \tref^{J} \}\,.
\end{equation} 
We define the subset of common parameters per given hypothesis as 
\begin{equation}\label{eq:common-pars}
 \bar{\params}_{\rm gw/kn/J} := \{D_L^{\rm gw/kn/J}, \iota^{\rm gw/kn/J}, \tref^{\rm gw/kn/J} \}
\end{equation} 
Further, by defining:
\begin{equation}
\bar{\bar{\params}}_{\rm gw} := \params_{\rm gw} / \bar{\params}_{\rm gw} \, , \quad \bar{\bar{\params}}_{\rm kn} := \params_{\rm kn} / \bar{\params}_{\rm kn} \, ,
\end{equation} 
with ``$/$'' standing for set subtraction, the set of parameters not shared among the two observations is then:
\begin{equation}
\bar{\bar{\params}}_{\rm J} = \bar{\bar{\params}}_{\rm gw} \cup \bar{\bar{\params}}_{\rm kn} \,,
\end{equation}
For concreteness, note that:
\begin{equation}
\bar{\bar{\params}}_{\rm J} = \{ m_1,m_2, \chi_1,\chi_2,\Lambda_1,\Lambda_2, \psi, \pref, \mej^{\dyn}, v^{\dyn}, 
  \kappa^{\dyn}, \mej^{\wind}, v^{\wind}, \kappa^{\wind} \}\,,
\end{equation} 
and that:
\begin{equation}
\params_{\rm J} = \bar{\params}_{\rm J} \cup \bar{\bar{\params}}_{\rm J}
\end{equation}
The prior distribution on the joint parameters can thus be derived as:
\begin{equation}
\label{eq:mm-prior-sp}
\pr(\params_{\rm J} | {\rm SS}) = 
\int \pr(\bar{\params}_{\rm gw}, \bar{\params}_{\rm kn}, \bar{\params}_{\rm J}, \bar{\bar{\params}}_{\rm J}) \, {\rm d} \bar{\params}_{\rm gw} {\rm d} \bar{\params}_{\rm kn} \, .
\end{equation}
Since the parameters not shared among the sources are independent of the shared ones, this simplifies to:
\begin{equation}
\pr(\params_{\rm J} | {\rm SS}) = 
\pr(\bar{\bar{\params}}_{\rm J}) \int \pr(\bar{\params}_{\rm gw}, \bar{\params}_{\rm kn}, \bar{\params}_{\rm J}) \, {\rm d} \bar{\params}_{\rm gw} {\rm d} \bar{\params}_{\rm kn} \, .
\end{equation}
Now we can impose the common source hypothesis, which implies:
\begin{equation}\label{eq:SS}
\pr(\params_{\rm J} | {\rm SS}) = 
\pr(\bar{\bar{\params}}_{\rm J}) \int \pr(\bar{\params}_{\rm gw}) \, \delta(\bar\params_{\rm gw} - \bar\params_{\rm J}) \, \delta(\bar\params_{\rm kn} - \bar\params_{\rm J}) \, {\rm d} \bar{\params}_{\rm gw} {\rm d} \bar{\params}_{\rm kn} \, ,
\end{equation}
where $\delta(x)$ is the Dirac distribution.
This immediately yields:
\begin{equation}
\pr(\params_{\rm J} | {\rm SS}) = \pr(\bar{\params}_{\rm J}) \pr(\bar{\bar{\params}}_{\rm J}).
\end{equation}
The case discussed above enforces a ``minimal connection'' for a single source,
and is weakly dependent on the specific models employed to describe the data. 
In the following, as $\pr(\bar{\params}_{\rm J})$ we use a volumetric prior on the inclination $\iota^{J}$ and luminosity distance $D_L^{J}$, 
and a uniform prior on $\tref^{J}$, with boundaries large enough to encompass the full posterior mass. 
For ease of notation, below we are going to drop the $J$ superscript on the signals parameters.

Further correlation among instrinsic parameters may be introduced
assuming a particular source model. 
For example, NR simulations can
provide phenomenological relations between the binary parameters and
the mass ejecta properties~\citep[\eg][]{Radice:2018ozg}.
The relations considered here specifically relate the dynamical ejecta
mass and velocity and the wind's mass 
$\params_{\rm ej}=\{\mej^{\dyn}, v^{\dyn},\mej^{\wind}\}$
to the binary masses and the tidal polarizability parameters,  
\begin{equation}
  \label{eq:fit-map}
  \params_{\rm ej}\mapsto \params_{\rm ej}=\nrmap(m_1,m_2,\Lambda_1,\Lambda_2)\,.
\end{equation}
By assuming Eq.~\eqref{eq:fit-map}, with the same formalism as above, we remove the dependency on
$\params_{\rm ej}$ and extend the set of common parameters to 
$\bar{\params}_{J} = \{m_{1, 2}, \Lambda_{1, 2}, D_L, \iota, \tref\}$.
Appendix~\ref{app:nr-map} describes in detail the construction of
$\nrmap$ used in this work. We comment here on two aspects.
First, we fit data from a large and etherogenuous set of NR
simulations. 
This provides only a conservative model and a proxy for the 
current systematic uncertainties on affecting ejecta and
kN light curves, \eg~\citep{Radice:2021jtw,Zhu:2020eyk,Barnes:2020nfi,Zappa:2022rpd}.
For the same reason, we do not consider information
on the average electron fraction, that could in principle also be
folded into the analysis \citep{Breschi:2021tbm}.
Second, NR relations carry errors of the order ${\sim}20\%$
that need to be taken into account during PE.
Within our Bayesian approach, we account for these uncertainties by
including auxiliary recalibration parameters and marginalizing over
these additional degrees of freedom~\citep{Breschi:2021tbm,Breschi:2021xrx}. 

Given a prior on the employed parameters,
the missing ingredient to define a Bayesian model
is the likelihood function.
Assuming that the GW and the kN observations are 
statistically uncorrelated,
we write the joint likelihood as
the product 
\begin{equation}
  \label{eq:mm-like}
  \pr(\data_{\rm gw},\data_{\rm kn}|\params_{\rm gw},\params_{\rm kn}) = 
  \pr(\data_{\rm gw}|\params_{\rm gw})\,
  \pr(\data_{\rm kn}|\params_{\rm kn})\,.
\end{equation}
Note that the analytical form of Eq.~\eqref{eq:mm-like}
does not depend on the prior choices.
While the case discussed here is specific for GW-kN, {\bajes{}}
implements a general framework for multimessenger datasets and reated
analyses.

\section{Results}
\label{sec:res}

\begin{table*}
  \centering    
  \caption{Median and 90\% credible interval of relevant parameters for the different analysis configurations.}
  \begin{tabular}{cccccc}        
    \hline\hline
    				 	 &  							  & GW170817 					& AT2017gfo			        & GW170817+AT2017gfo &  GW170817+AT2017gfo (NRI) \\
    \hline
    $M$    		     & $[\Mo]$ 		 			  & $2.9^{+ 0.3}_{-0.1}$         & - 						& $2.8^{+ 0.1}_{-0.1}$		  & 	$2.77^{+ 0.03}_{-0.01}$      \\
    $\M_c$           & $[\Mo]$ 		 			  & $1.1976^{+0.0005}_{-0.0002}$ & - 				        & $1.1976^{+0.0002}_{-0.0001}$ & $1.1975^{+0.0001}_{-0.0001}$ \\
    $q$              & 		   		 			  & $1.7^{+0.9}_{-0.6}$ 		    & - 						    & $1.4^{+0.3}_{-0.3}$ 		  & 	$1.2^{+0.2}_{-0.2}$          \\
    $\chi_{\rm eff}$ & 	 	   		 			  & $0.05^{+0.08}_{-0.05}$ 	    & - 						    & $0.02^{+0.03}_{-0.02}$		  & 	$0.00^{+0.01}_{-0.01}$       \\
    $\lambdat$ 	     & 		   		 		 	  & $160^{+485}_{-146}$ 			& - 						    & $684^{+540}_{-382}$		  & 	$667^{+382}_{-270}$	         \\
    $\delta\Lambda$  & 		   	    				  & $7^{+118}_{-84}$ 			    & - 						    & $37^{+185}_{-284}$			  & 	$266^{+157}_{-109}$          \\
    \hline
    $\mej^{\dyn}$    & $\times10^{-2}[\Mo]$ 		  & - 							& $8^{+3}_{-3}$		        & $0.1^{+0.1}_{-0.3}$	  & 		-				   		 	 \\ 
    $v^{\dyn}$ 		 & $c$ 			 	   		  & - 					        & $0.30^{+0.03} _{-0.06}$    & $0.30^{+0.02} _{-0.02}$ & 		-				   			 \\ 
    $\kappa^{\dyn}$  & $[{\rm cm}^2~{\rm g}^{-1}]$ & - 						    & $0.16^{+0.04}_{-0.02}$     & $23^{+8}_{-13}$		 & $0.20^{+0.04}_{-0.04}$  			 \\ 
    $\mej^{\wind}$   & $\times10^{-2}[\Mo]$ 		  & - 							& $10^{+3}_{-3}$ 	         & $12^{+3}_{-2}$         & 	-					   			 \\ 
    $v^{\wind}$      & $c$ 						  & - 							& $0.046^{+0.007}_{-0.015}$  & $0.01^{+0.03}_{-0.01}$ & $0.03^{+0.01}_{-0.01}$  			 \\ 
    $\kappa^{\wind}$ & $[{\rm cm}^2~{\rm g}^{-1}]$ & - 							& $0.5^{+0.1}_{-0.2}$        & $0.02^{+0.08}_{-0.02}$	 & $0.4^{+0.2}_{-0.1}$ 				 \\ 
    \hline
    $D_L$ 			 & $[\Mpc]$ 		 			  & $36^{+12}_{-13}$ 		    & $45^{+5}_{-8}$ 			& $42^{+2}_{-3}$			      & 	$36^{+5}_{-4}$                \\
    $\iota$ 		     & $[{\rm rad}]$ 			  & $2.4^{+0.5}_{- 0.4}$ 		& $1.5^{+1}_{-1}$           & $2.7^{+0.1}_{-0.1}$	      & 	$2.4^{+0.2}_{-0.2}$           	 \\
    \hline
    $\log(L_{\rm max})$&							  &	528.92					    & -158.22                   & 286.69					    & 366.52				     	  	  \\
    \hline\hline
  \end{tabular}
 \label{tab:ci}
\end{table*}

\begin{figure*}[h!]
  \centering 
  \includegraphics[width=0.98\textwidth]{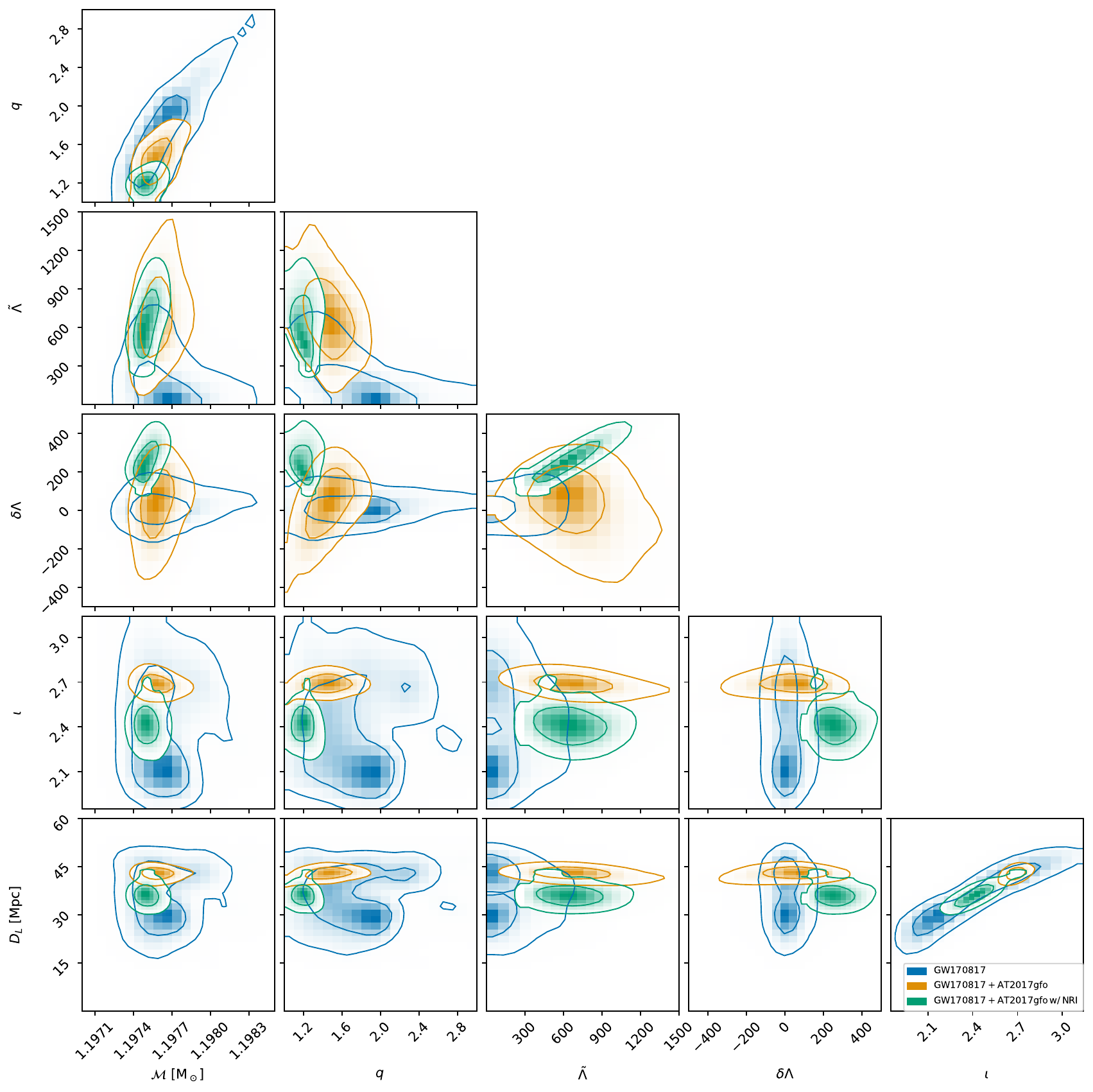}
  \caption{Posterior distribution for GW relevant parameters under the various assumptions.
    The contours report the 50\% and the 90\% credibility regions.
    Blue posteriors are computed from GW170817 data only, orange
    (green) posteriors correspond to the joint GW-kN data without
    (with) NR-informed mappings.} 
  \label{fig:pos_gw}
\end{figure*}   

\begin{figure*}[h!]
  \centering 
  \includegraphics[width=0.98\textwidth]{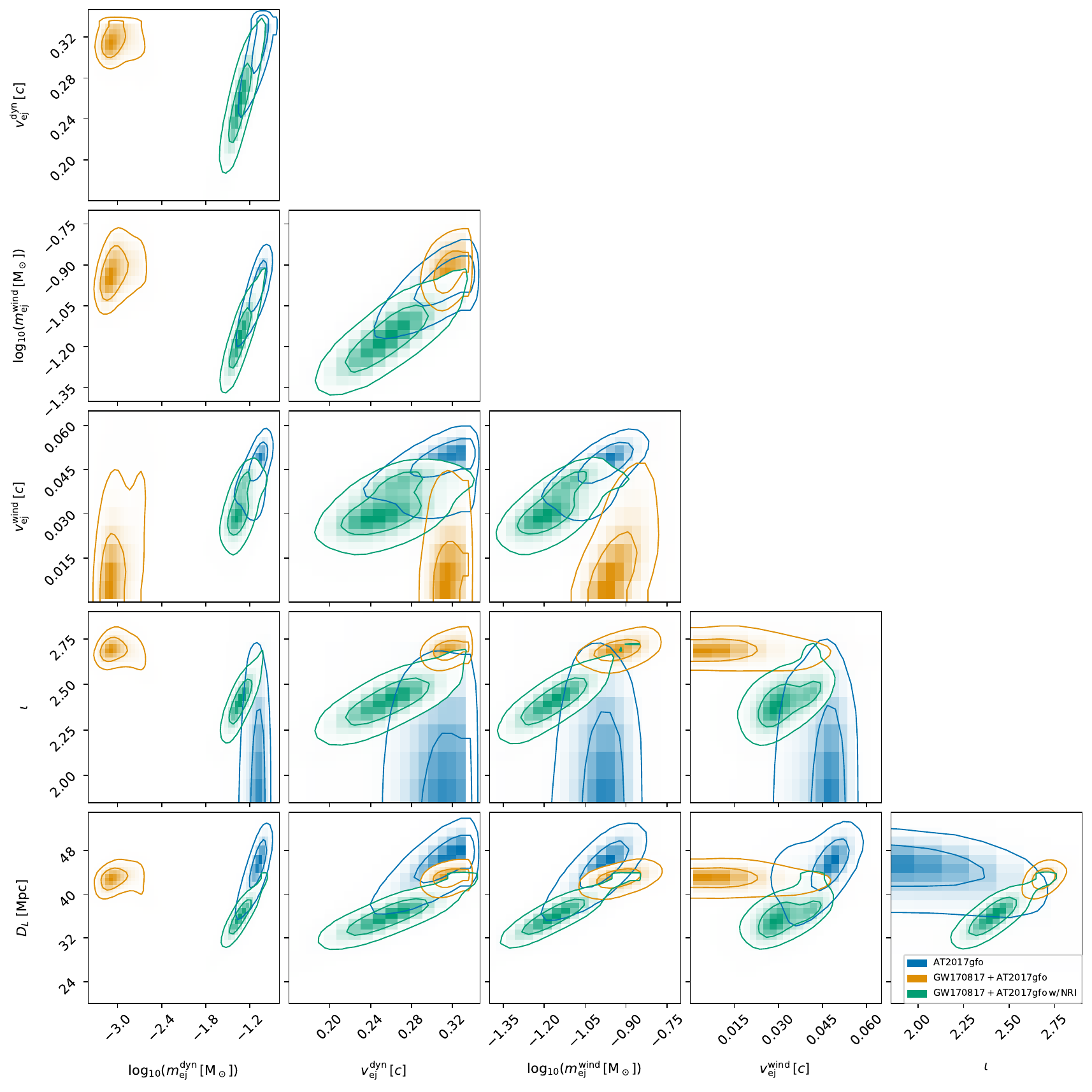}
  \caption{Posterior distribution for kN relevant parameters under the various assumptions.
    The contours report the 50\% and the 90\% credibility regions.
    Blue posteriors are computed from AT2017gfo data
    only, orange (green) posteriors correspond to the
    joint GW-kN data without (with) NR-informed
    mappings.}
  \label{fig:pos_kn}
\end{figure*}     

We first perform PE on GW170817 and AT2017gfo assuming the two
signals to have originated from different sources, and then two joint
analyses: one assuming a ``minimally informed'' prior and another
using the NR-informed prior. 
The results of our analyses are summarized in Tab.~\ref{tab:ci} and Figs.~\ref{fig:pos_gw}-\ref{fig:pos_kn}.  

\subsection{Single messenger analyses}
\label{sec:res:sm}

We start by briefly discussing the inference on GW and kN signals,
when analysing each dataset separately. 

For the GW170817 analyses, we obtain a binary (detector-frame) mass of
$M=2.9^{+0.3}_{-0.2}~\Mo$ and a mass ratio of $q=1.7^{+0.9}_{-0.6}$,
while the measurement of the reduced tidal parameter yields $\lambdat
= 160^{+485}_{-146}$.  
The luminosity distance is constrained to $D_L=36^{+12}_{-13}~{\rm Mpc}$.
These results are consistent with previous analyses~\citep[see,
  \eg][]{Abbott:2018wiz,Breschi:2021wzr,Tissino:2022thn}.
Note that the mass ratio can be sensitive to sampling errors, in
particular showing tail extending to large values~\citep{Tissino:2022thn}.

For the AT2017gfo analysis, we measure $\mej^{\dyn}=8^{+3}_{-3}{\times
}10^{-2}~\Mo$ and $v^{\dyn}=0.30^{+0.02}_{-0.04}$~c, for the mass and
velocity of the first component. 
For the second component, we obtain
$\mej^{\wind}=1.0^{+0.2}_{-0.3}{\times }10^{-1}~\Mo$ and
$v^{\wind}=4.6^{+0.6}_{-1.4}{\times }10^{-2}$~c.  
These ejected masses and velocities are broadly consistent with
previous estimates using spherically symmetric
models~\citep[\eg][]{Cowperthwaite:2017dyu,Villar:2017wcc,Coughlin:2018miv,Breschi:2021tbm}. 
%
%%\dr{These masses seem way to large to me.}  
Despite our agnostic prior choice, the PE points towards one component
to be less massive and signficantly faster with respect to the second
component. 
The inferred ejecta velocities for the lighter (heavier) and faster
(slower) components are compatible with the average values predicted
by NR simulations for the dynamical and wind
ejecta~\citep{Nedora:2020hxc}, with the heavier component interpreted
as a massive wind emerging from the delayed collapse of the remnant
neutron star~\citep{Radice:2018pdn,Nedora:2019jhl,Kiuchi:2022nin,Radice:2023xxn}.
However, the inferred dynamical ejecta mass overestimates current NR
results by about one order of magnitude.  
This appears to be a common feature in most of the analyses performed
so far, and indicates the need for more sophisticated kN models.  

The gray opacities inferred values are more difficult to interpret.
On the PE side, this parameter is highly degenerate with distance
inclination and total mass, since it controls the signal luminosity.  
On the modeling side, it is known that the complex atomic and
radiation transport physics cannot be adequately captured by a single,
averaged and time independent parameter, see 
\eg~\citep{Ricigliano:2023svx} for a recent discussion.
The luminosity distance is constrained to $D_L=45^{+5}_{-8}~{\rm
  Mpc}$, consistent with GW estimates and measurements of the host
galaxy NGC~4993.  

\begin{figure*} 
  \centering 
  \includegraphics[width=0.98\textwidth]{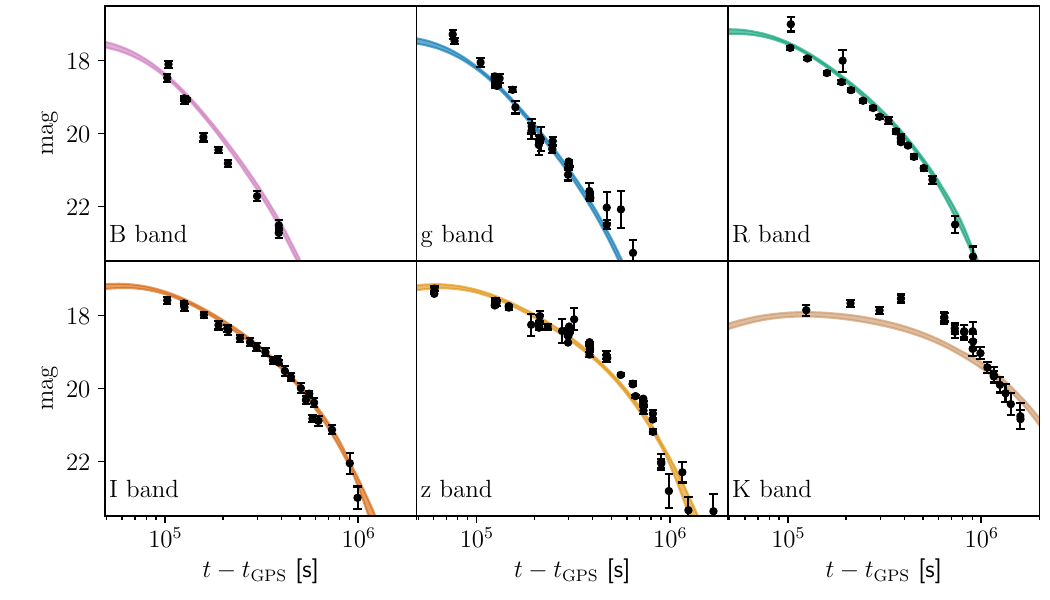}
  \caption{Light curve reconstruction from the inference on
    AT2017gfo. The bands represent the parameter variation within
    their 90\% credible regions.} 
  \label{fig:LC}
\end{figure*}   

In Fig.~\ref{fig:LC} we show the reconstructed light curves.
The employed kN model allows to capture the bulk of the data trend. 
The quantitative behaviour of the light curve is reproduced over the
${\sim}20$~days period analyzed for several photometric bands.  
An exception is the pre-peak behaviour in the
$K$-band, which is not accurately reproduced.
Further investigations are needed to interpret this feature.
Comparing to our previous work \citep{Breschi:2021tbm}, the light
curves and opacities were better captured with the parameters and
prior choices made there~\footnote{
Note that also a different sampler from \cite{cpnest2} was utilized in that work.
}.
We stress that the spherical model used here is not the best-fitting semi-analytical kN
model for these analyses, but it is employed
here as a first step towards more complex inferences. 

\subsection{Joint \& coherent analyses}
\label{sec:res:mm}

We now discuss the impact of the joint analysis on the posterior distribution of the two sets of parameters when performing a joint and coherent analysis.

\subsubsection{Agnostic prior on intrinsic parameters}

For this analysis, we assume a common source but no additional NR-informed priors.

Concerning the GW parameters, a noticable improvement is observed on the distance (and indirectly on the inclination parameter, as discussed below), now shared among the two models, with a significant tightening of the posterior volume.
The chirp mass of the binary remains consistent with the GW-only measurement, with its error bars shrinking.
This is not due to direct information on the parameter itself, but is an indirect consequence of the shrinkage in the mass ratio posterior, whose tail is significantly cut by the additional information, pointing towards a comparable-mass system.
The improvement in the mass ratio measurement can be in turn traced to the better measurement of the source distance, due to the impact that these parameters share on the GW signal amplitude.
The system is now located further away, an effect which can in part be compensated by a higher GW intrinsic luminosity obtained for a more equal-mass system, or by changing the inclination angle.
Similarly, the correlation between $q$ and $\tilde{\Lambda}$ implies that larger values of the effective tidal parameter are now favoured.
Instead, $\delta \Lambda$ is weakly affected and remains consistent with zero.

Concerning the kN parameters, the velocity of the dynamical component remains essentially unaffected by the joint analysis.
The dynamical ejecta mass instead shifts towards much smaller values. 
This effect is again due to the reduced distance value, which increases the intrinsic source luminosity, hence requiring a lower amount of ejecta to remain consistent with the data.
For the wind parameters, the mass remains consistent with the single-source analysis, while the velocity and opacity are pushed towards smaller values, albeit attaining values within the same order of magnitudes with respect to the previous analysis.

\subsubsection{NR-informed prior on intrinsic parameters}

For this analysis, we assume a common source and the
additional NR-informed prior descibed above, thus constraining a subset of
the kN intrinsic parameters through binary ones. 

Concerning the GW parameters, the chirp mass of the binary keeps
remaining consistent with the GW-only measurement, with its error bars
slightly shrinking again. 
This happens for the same reason explained above, as it can be
appreciated by the overlap between the green posterior in the top left
panel of Fig.~\ref{fig:pos_gw} with the blue posterior. 
The mass ratio is further shrinked by the NR information, becoming
even more consistent with unity, with an impact on $\tilde{\Lambda}$
similar to the one discussed in the case of an agnostic analysis. 
The inclination remains consistent with the joint analysis performed
under the agnostic prior, albeit with a smaller error bar due to a better 
measurement of the mass ratio, as discussed in the previous case. The NR
information also helps in breaking of the degeneracy of the mass ratio with the distance
(bottom rightmost panel of Fig.~\ref{fig:pos_gw}). 
The largest impact from the NR information is imparted on $\delta
\Lambda$. While still consistent with the previous measurement, its
posterior now prefers non-zero values. 

Concerning the kN parameters, the source is now inferred as closer and
more off-axis, pushing the wind mass to smaller values. The mass of
the dynamical component becomes again consistent with the
single-source analysis. The velocities are only mildly affected by the
NR-informed prior, with the wind velocity moving away from zero
towards values which are again consistent with the kN-only analysis. 

\subsubsection{Bayes' factors}
\label{sec:res:bf}

Within a Bayesian framework, and exploiting the nested sampling
algorithm we employed above, it is immediate to compare the evidence
for different hypothesis explaining the dataset under consideration,
when marginalising over the whole parameter space.
Specifically, we are interested in comparing two hypotheses: the
``coherent'' one, where GW and kN data are simultaneously modelled by
a single common source, which we refer to as ``SS'' following
Eq.~\eqref{eq:SS}, and the ``incoherent'' hypothesis, where the two
datasets are each explained by independent and disjoint sources, which
we refer to as ``DS'' following Eq.~\eqref{eq:DS}.  

The BF comparing these hypotheses can be obtained through the
coherence ratio introduced in~\cite{Veitch:2009hd}: 
\begin{equation}
  \label{eq:bf}
  \B^{\rm SS}_{\rm DS} = \frac{\int d\theta_J p(\theta_J | \rm SS)
    \pr(\data_{\rm gw}|\theta_J, \rm SS) \pr(\data_{\rm kn}|\theta_J,
    \rm SS)}{\int d\theta_{\rm gw} p(\theta_{\rm gw} | \rm SS)
    \pr(\data_{\rm gw}|\theta_{\rm gw}, \rm SS) \int d\theta_{\rm kn}
    p(\theta_{\rm kn} | \rm DS) \pr(\data_{\rm kn}|\theta_{\rm kn},
    \rm DS)}\,. 
\end{equation}
As discussed in~\cite{Veitch:2009hd} in more detail, this ratio
intuitively compares the integral of the product with the product of
the integrals of the distributions, a measure of how much information
is gained by assuming a joint hypothesis. 
Additionally, we label by ``SS-NR'' the hypothesis obtained when
considering the NR-calibrated relation Eq.~(\ref{eq:fit-map}). 
When applying this computation to the above results, we find:
\begin{equation}
  \label{eq:bf_SS}
  \log_e(\B^{\rm SS}_{\rm DS}) = -44.23 \pm 0.2 \, ,
\end{equation}
without assuming NR relations, and:
\begin{equation}
  \label{eq:bf_SS-NR}
  \log_e(\B^{\rm SS-NR}_{\rm DS}) = 46.49 \pm 0.2 \, ,
\end{equation}
when assuming NR-calibrated relations.

These results indicate that, within the available dataset and under
the employed models, the two observations are better explained by the
common source hypothesis (SS) only assuming NR-informed relations.
An incoherent explanation (DS) is instead favoured when assuming that
only the distance, inclination and coalescence time are 
common parameters. 
Note that, since the kN model is spherical, the key
parameter is the distance, but the latter is
compatibile among the single messenger analyses.
Thus, the negative value of $\log_e(\B^{\rm SS}_{\rm DS})$ simply
reflects the fact that the distance is consistent in both single
messenger analyses and a coherent analysis does not help in fitting the data
more than the single messenger analyses.
Overall, this result is related to the simplified kN model used
in this work and the high degeneracy in kN parameter space discussed above.
This underlines the relevance of systematics in kN modeling, and
enforces the importance of informing the models with full numerical
solutions by connecting them with binary parameters.

\section{EOS constraints}
\label{sec:eos}

In this section we discuss the constraints on the EOS from our
analysis. We follow an approach similar to the one presented in
\citep{Breschi:2021tbm,Breschi:2021xrx}, with a few key
differences. First, we use the updated EOS-insentive NR 
relations developed here. Second, we compute a new set of ${\sim} 10$~million
phenomenological EOS by employing minimal assumptions \cite[]{Godzieba:2020tjn}.
The set is generated with a Markov Chain Monte Carlo approach by
fixing the crust EOS and assuming only i) general relativity; ii) 
causality at higher densities. Hence, it is agnostic on nuclear physics and not affected by nuclear
physics uncertainties.   
The set includes EOS with and without first order phase transitions.
Differently from \cite[]{Godzieba:2020tjn}, we do not include any
constraints from GW170817 and we sample EOS such that $\Mmax>2.09\Mo$ \citep{Romani:2022jhd}
a posteriori, as described below.
Third, we fold-in the pulsars results from~\cite{Miller:2019cac,Vinciguerra:2023qxq}. 
In order to consistently account for the information listed above we
sample over the
EOS index in our $10$~million set $i_{\rm EOS}$, as well as over the masses 
of PSR J0030+0451 ($M_{\rm J0030}$), 
PSR J0740+6620 ($M_{\rm J0740}$) and the
  masses of the two components of GW170817 progenitor  
system ($m_1, m_2$) \citep[]{Foreman-Mackey:2013}.
Denoting $ \data_{\rm MM} = \{m_1, m_2, \tilde{\Lambda}\}$, 
            $ \data_{\rm J0030} = \{M_{\rm J0030}, R_{\rm J0030}\}$
        and $ \data_{\rm J0740} = \{M_{\rm J0740}, R_{\rm J0740}\}$,
the likelihood we employ is given by:
\begin{multline}
p ( \data | i_{\rm EOS}, m_1, m_2, M_{\rm J0030}, M_{\rm J0740})  = p_{\rm MM}(\data_{\rm MM}| i_{\rm EOS}, m_1, m_2) \\
\times p_{\rm J0030}(\data_{\rm J0030}| i_{\rm EOS}, M_{\rm J0030} ) \times p_{\rm J0740}(\data_{J0740} | i_{\rm EOS}, M_{\rm J0740} )  \, ,
\end{multline}
where $p_{\rm MM}(\dots)$, $p_{\rm J0030}(\dots)$, $p_{\rm J0740}(\dots)$ 
are obtained as a gaussian KDE of posteriors from the respective analyses.
Similarly, the priors on $M_{\rm J0740}, M_{\rm J0030}, m_1, m_2$ are obtained from 
their marginalized one-dimensional posterior, while the prior on the
EOS index is assumed to be uniform. 

\begin{figure*} 
  \centering 
  \includegraphics[width=0.98\textwidth]{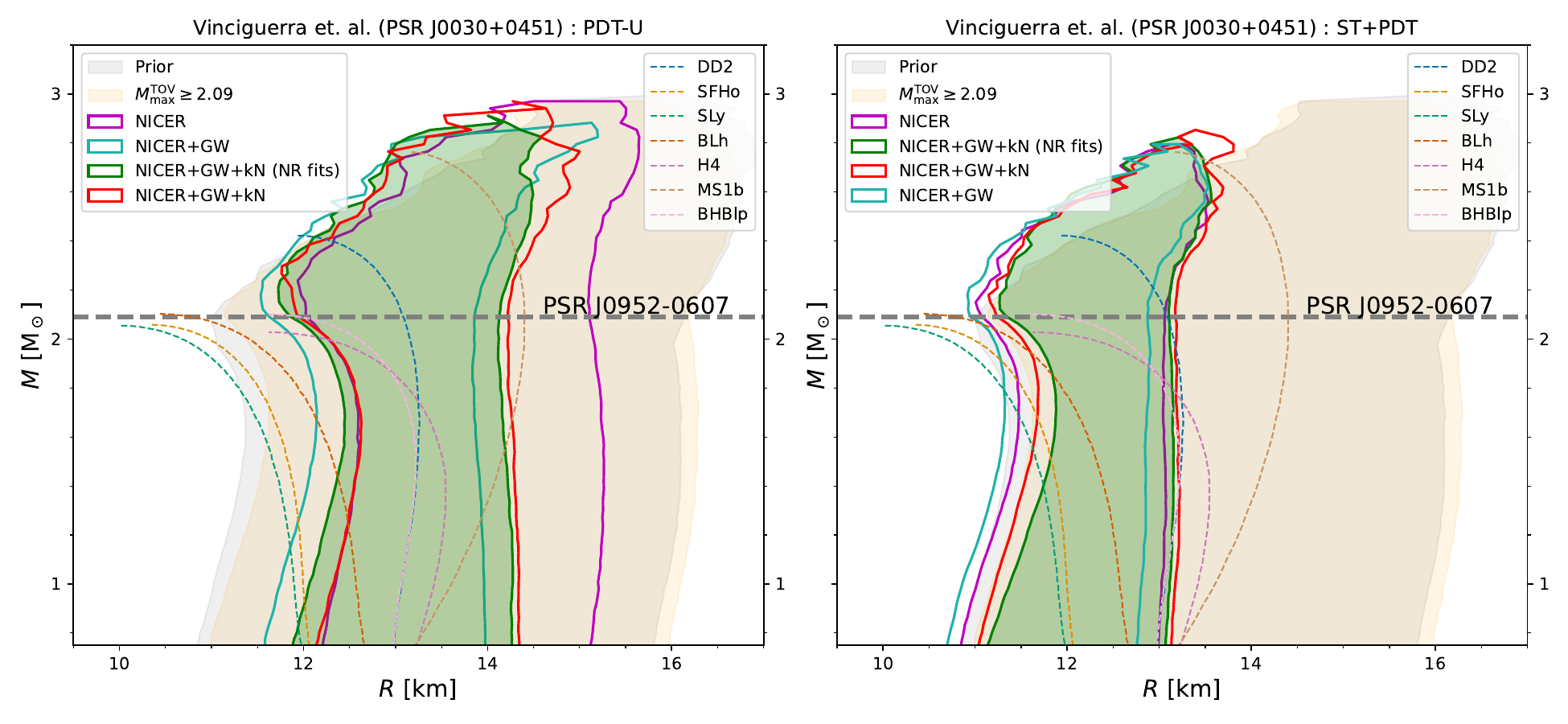}
  \caption{Constraints on the EOS obtained by resampling the posteriors of the GW-only (blue), 
  joint (red) and NR-informed joint analysis (green) using 10 million EOS and
  folding in the NICER information of \citep{Vinciguerra:2023qxq},
  \citep{Miller:2021qha} and the measurement of $\Mmax >
  2.09\Mo$~\citep{Romani:2022jhd}. 
  Depending on the hot-spot model employed for the analysis of PSR J0030+0451
  (PDT-U or ST+PDT, respectively shown in the left and right panels), 
  the EOS constraints shift by ${\sim} 1$~km, indicating that (i) NICER data provides 
  strong constraints on the EOS, and (ii) the systematic errors in such analyses are large.}
  \label{fig:eos_constraints}
\end{figure*}

Results are displayed in Fig.~\ref{fig:eos_constraints}, and
summarized in Tab~\ref{tab:eos_constraints}. 
The addition of kN data to the GW analysis results in a systematic
shift of the $\R14$ posterior to larger radii of ${\sim}0.5$~km. This
is essentially related to the larger $\tLam$ median and, physically,
to the fact that the kN places a lower bound to $\tLam$ due to disc
wind~\citep{Radice:2017lry}. 
The addition of the NR-informed relations to the inference has a
relatively small impact on the  
EOS constraints, with the estimated values of $\R14$ and
$\Lambda_{1.4}$ compatible within the error bars 
with the same quantities from the joint analysis.
This result is not surprising, as the recovered $\tilde\Lambda$ and
$\mathcal{M}$ distributions of the two analyses  
are largely consistent (see Fig.~\ref{fig:pos_gw}).

The impact of NICER data on the allowed EOSs, instead, is
substantial. The permissibility of certain EOSs (\eg~MS1b) depends on
the specific hot-spots model employed (PDT-U or ST+PDT). The estimated  
distributions of $\R14$ are shifted of ${\sim} 1$~km between the two
models, indicating large systematic errors 
in the NICER analyses~\citep{Vinciguerra:2023qxq}.
Notably, while the evidence of the NICER analysis favors the PDT-U
model, the hot spot configuration predicted by the ST+PDT model was
found  
to be more consistent with the gamma-ray emission associated with PSR
J0030+0451~\citep{Kalapotharakos:2020rmz} .

\begin{table*}[t]
  \centering    
  \caption{EOS constraints obtained from our multi-messenger PEs.
  Starting from our agnostic prior set of 10 million EOS, 
  we progressively add the information coming from PSR J0952+0607 ($\Mmax > 2.09\Mo$),
  PSR J0030+0451 and PSR J0740+6620 (NICER), GW170817 (GW) and AT2017gfo (kN).
  We report the median and $90\%$ credible intervals of the relevant EOS-dependent properites.}
  \begin{tabular}{lcccccc}        
    \hline\hline
    Data & $\R14$ & $\Lambda_{1.4}$ & $\Mmax$ & $\Rmax$ & $\log_{10}P(2\rhosat)$ & $\log_{10}P(4\rhosat)$ \\
         & [km]   &                & [$\Mo$] & [km]    & & \\
    \hline
    Prior                                       & $ 12.95^{+ 3.10}_{- 1.65}$& 
                                                  $ 540^{+ 1410}_{- 310}$ & 
                                                  $ 2.27^{+ 0.52}_{- 0.26}$ & 
                                                  $ 11.90 ^{+ 2.80}_{- 1.54 }$ & 
                                                  $ 34.72^{+ 0.35}_{- 0.29}$ & 
                                                  $ 35.50^{+ 0.27}_{- 0.18}$\\
    $\Mmax$       &                             $ 13.23^{+ 2.98}_{- 1.73}$ & 
                                                $ 620 ^{+ 1470}_{- 350}$ &
                                                $ 2.32^{+ 0.50}_{- 0.22}$ & 
                                                $ 12.15^{+ 2.72}_{- 1.62}$ &
                                                $ 34.76^{+ 0.32}_{- 0.29}$ &
                                                $ 35.54^{+ 0.26}_{- 0.20}$ \\
    \hline
    $\Mmax$+NICER$^{\rm ST+PDT}$                & $ 12.08^{+ 0.93}_{- 0.7}$& 
                                               $ 360^{+ 190}_{- 110}$ & 
                                               $ 2.25^{+ 0.26}_{- 0.15}$& 
                                               $ 11.30^{+ 1.32}_{- 0.82}$ & 
                                               $ 34.61^{+ 0.18}_{- 0.18}$& 
                                               $ 35.53^{+ 0.14}_{- 0.12}$\\
    $\Mmax$+NICER$^{\rm ST+PDT}$+GW            & $ 11.86^{+ 0.95}_{- 0.66}$ & 
                                                 $ 330^{+ 180}_{- 100}$ & 
                                                 $ 2.25^{+ 0.23}_{- 0.14}$ &
                                                 $ 11.15 ^{+ 1.23 }_{- 0.74}$ & 
                                                 $ 34.56 ^{+ 0.18 }_{- 0.17 }$ & 
                                                 $ 35.53 ^{+ 0.13 }_{- 0.10 }$ \\
    $\Mmax$+NICER$^{\rm ST+PDT}$+GW+kN            & $ 12.33^{+ 0.84}_{- 0.81}$ & 
                                                    $ 410^{+ 180}_{- 130}$& 
                                                    $ 2.27^{+ 0.25}_{-0.15}$ &
                                                    $ 11.48^{+ 1.30}_{- 0.90}$   & 
                                                    $ 34.65^{+ 0.16}_{- 0.17}$&
                                                    $ 35.54^{+ 0.14}_{-0.12}$ \\
    $\Mmax$+NICER$^{\rm ST+PDT}$+GW+kN (NRI)  & $ 12.30^{+ 0.81}_{- 0.56}$ & 
                                                    $ 400^{+ 180}_{-100}$ & 
                                                    $ 2.28^{+ 0.25}_{- 0.16}$& 
                                                    $ 11.53^{+ 1.15}_{- 0.90}$& 
                                                    $ 34.64^{+ 0.16}_{- 0.12}$    &
                                                    $ 35.54^{+ 0.14}_{- 0.12}$ \\
    \hline
    $\Mmax$+NICER$^{\rm PDT-U}$                  & $ 13.64^{+ 1.60 }_{- 1.18}$ &
                                                   $ 740^{+ 670}_{- 300}$ & 
                                                   $ 2.35^{+ 0.40}_{- 0.24 }$ &
                                                   $ 12.54^{+ 1.77}_{- 1.45 }$ & 
                                                   $ 34.61^{+ 0.18}_{- 0.18}$ & 
                                                   $ 35.53^{+ 0.14}_{- 0.12 }$ \\
    $\Mmax$+NICER$^{\rm PDT-U}$+GW               & $ 12.86 ^{+ 0.90 }_{- 0.83 }$ & 
                                                   $ 540 ^{+ 230}_{- 190 }$ & 
                                                   $ 2.30 ^{+ 0.31 }_{- 0.17 }$& 
                                                   $ 12.08 ^{+ 1.54 }_{- 1.24 }$& 
                                                   $ 34.74 ^{+ 0.20 }_{- 0.15 }$ & 
                                                   $ 35.53 ^{+ 0.18}_{- 0.21 }$\\
    $\Mmax$+NICER$^{\rm PDT-U}$+GW+kN            & $ 13.30^{+ 0.87}_{- 0.81}$ & 
                                             $ 640^{+ 280}_{- 190}$    &
                                             $ 2.30^{+ 0.30}_{- 0.20}$    & 
                                             $ 12.33^{+ 1.55}_{- 1.28}$ & 
                                             $ 34.77^{+ 0.20}_{- 0.11}$    & 
                                             $ 35.54^{+ 0.17}_{- 0.22}$\\
    $\Mmax$+NICER$^{\rm PDT-U}$+GW+kN (NRI) & $ 13.20^{+ 0.91}_{- 0.90}$ & 
                                             $ 620^{+ 270}_{- 200}$ &
                                             $ 2.32^{+ 0.30}_{- 0.19}$ & 
                                             $ 12.21^{+ 1.61}_{- 1.24}$ &
                                             $ 34.76^{+ 0.21}_{- 0.12}$ &
                                             $ 35.54^{+ 0.17}_{- 0.22}$ \\
    \hline\hline
  \end{tabular}
 \label{tab:eos_constraints}
\end{table*}

\section{Conclusions}
\label{sec:conclusion}

We introduced \bajes{-MMA}, a multi-messenger Bayesian pipeline for 
analyses of signals from binary neutron star mergers.
The key features of \bajes{} are:
(i) The use of NR relations for the ejecta mass and velocities; 
(ii) The use (and concurrent marginalization) of suitable
recalibration parameters, which widen the credible ranges of inferred
parameters accounting for systematic uncertainties;
(iii) The possibilities of performing joint and coherent analyses.
Regarding (ii),  this technique is here applied to handle the
systematics of phenomenological NR relations used to link remnant
properties with the binary properties.
More generally, the same methodology could be employed to deal
with other types of uncertainties, \eg~in the EOS set.
We note that the open source \bajes{-MMA} implementation is generic
and can accomodate different likelihood, datasets and models. 

Our inference results confirm previous findings on the presence 
of multiple kN components: a faster and lighter ``dynamical'' component, 
and a slower and heavier ``wind'' component,
\eg~\citep{Villar:2017wcc,Perego:2017wtu,Breschi:2021tbm}. 
They also indicate consistency in the source
distance parameter between GW and kN data.
Within spherical kN models, a common GW-kN source is favoured 
by our joint and coherent analysis only when assuming NR-informed 
relations between the ejecta components and the binary parameters, 
capable of significantly breaking the correlations in the GW-kN parameter space.
This result also highlights the impact of systematics in kN 
modeling, and that the degeneracies in kN parameter space can be 
effectively reduced by incorporating in the inference parametric
relations with binary parameters~\citep{Raaijmakers:2021slr}.

Multimessenger analyses with multiple sources can help improving
mass-radius diagram constraints. Current observations of GWs and PSRs
along with minimal EOS assumptions (validity of general relativity, causality)
point to NS maximum masses of $\Mmax\sim2.25-2.32$ with errors of
${\sim}11\%$ and NS radii $\R14\sim12-13$~km with ${\sim}1$~km
uncertainty and same order of magnitude systematics.
Our results indicate that current gravitational-wave
constraints are compatible with pulsars constraints, in-line with
other similar analyses.
Pulsars constraints however may be particularly sensitive to
systematics. For the data considered here this is clearly the case of
J0740+6620 \citep{Vinciguerra:2023qxq}. 
Our mass-radius constraints can be translated into
constraints into pressure-density constraints for the EOS. For
example, at twice saturation densities we find 
$\log P(2\rho_{\rm sat})\simeq34.61-34.76$ (depending on systematics), see
Tab.~\ref{tab:eos_constraints}. Interestingly, such a constraint
potentially excludes some of the EOS commonly employed in NR
simulations.
Future observations of BNSM might
improve the precision of these constraints and also help to break
systematics effects. 

  In order to compare our new inferences to previous work it is
  useful to focus on the NS radius, which is a commonly inferred parameter.
  Our results are in good agreement with some of the first
  inferences performed on this set of
  data~\eg~\citep{Radice:2018ozg,Coughlin:2018fis} (See Fig.~12 of 
  \cite{Breschi:2021tbm} for a collection of various results). On the
  one hand, we confirm the robusteness of the constraint in joint and
  coherent analysis. On the other hand, we point out out that part of
  the agreement is related to the dominant effect of the
  minimum-maximum mass constraint from pulsar data.
  Additionally, our work leverage on the recent pulsar analysis of 
  \cite{Vinciguerra:2023qxq} to show that the systematics
  in the radius measurement from pulsar data can be significantly
  larger than those from GW+kN.

Future work will be devoted to explore more sophisticated kN models
as those included in the \texttt{xkn} framework \citep{Ricigliano:2023svx},
as well as kN afterglow models \citep{Hajela:2020a,Nedora:2021eoj}.
As discussed in Sec.~\ref{sec:res:sm}, an main issue in the
  parameter estimation with analytical kN models 
  appear to be the interpretation of the effective gray opacity parameters.
  On the one hand, it is unrealistic to expect that such a parameter
  can capture the complexity of the atomic physics in kN
  \citep{Zhu:2020eyk, Barnes:2020nfi}. On the other hand, the opacity
  modelling is a significant issue also when numerical models are
  employed, \eg~\citep{Bulla:2022mwo}. Also in that case very simplified
  models are employed~\footnote{ %
  For example, the recent analytical model of \cite{Ricigliano:2023svx} applies
  the same prescription as in some numerical models
  \citep{Bulla:2022mwo}; and opactities can be either prescribed or inferred.
  },  but because they are not inferred, they remain
  often hidden in the model assumption. Overall, photon transport,
  hydrodynamical interaction and atomic/nuclear physics in kN modeling  
  remain a stand-alone key challenge (in large part independent on Bayesian PE.)

Improved NR relations can be created by utilizing a more homogeneous
set of microphysical simulations \citep{Nedora:2020qtd} and including the
contribution of the disc winds from upcoming long-term simulations
\cite{Kiuchi:2022nin,Radice:2023zlw}. This is another avenue we plan
to explore in the future.

We also plan to extend the \bajes{-MMA} framework to
include likelihood and models for GRB and afterglow data, \eg
\citep{Biscoveanu:2019bpy,Hayes:2019hso,Farah:2019tue,Gianfagna:2022kpw}.
The inclusion of this messenger can improve the inference of the 
extrinsic parameters of the source, in particular the viewing angle,
with implication for cosmological parameters.
However, the inclusion of GRB data in joint analyses is not expected
to provide valuable extra information on the EOS because the link
between the GRB emission and the NS matter properties is still unclear,
\eg~\citep{Piran:2012wd,Hotokezaka:2018aui} (see also
\cite{Farah:2019tue} for similar consideration on the jet structure.)

\begin{acknowledgements}
SB acknowledges hospitality and support from the IGC at PSU, where this
work was finalized. 

MB acknowledges support from the European Union’s H2020 under ERC
Starting Grant, no.~BinGraSp~714626; from the Deutsche
Forschungsgemeinschaft (DFG) under Grant no.~406116891 within the
Research Training Group (RTG) 2522/1; and from the PRO3 program
``DS4Astro'' of the Italian Ministry for Universities and Research. 
RG acknowledges support by the Deutsche Forschungsgemeinschaft (DFG) under Grant No.
406116891 within the Research Training Group RTG 2522/1 and from NSF Grant PHY-2020275
(Network for Neutrinos, Nuclear Astrophysics, and Symmetries (N3AS))
GC acknowledges funding from the Della Riccia Foundation under an
Early Career Scientist Fellowshis, from the European Union’s Horizon
2020 research and innovation program under the Marie Sklodowska-Curie
grant agreement No. 847523 ‘INTERACTIONS’, from the Villum
Investigator program supported by VILLUM FONDEN (grant no. 37766) and
from the DNRF Chair, by the Danish Research Foundation. 
SB acknowledges funding from the EU Horizon under ERC Consolidator
Grant, no. InspiReM-101043372 and from the Deutsche
Forschungsgemeinschaft, DFG, project MEMI number BE 6301/2-1.
DR acknowledges funding from the U.S. Department of Energy, Office of
Science, Division of Nuclear Physics under Award Number(s) DE-SC0021177,
DE-SC0024388, and from the National Science Foundation under Grants No.
PHY-2011725, PHY-2116686, and AST-2108467.
The computations were performed on the ARA cluster at Friedrich
Schiller University Jena, on the supercomputer SuperMUC-NG at the Leibniz-
Rechenzentrum (LRZ, \url{www.lrz.de}) Munich.
{\scshape ARA}, is a resource of Friedrich-Schiller-Universt\"at Jena
supported in part by DFG grants INST 275/334-1 FUGG, INST 275/363-1
FUGG and EU H2020 BinGraSp-714626. 
The authors acknowledge the Gauss Centre for Supercomputing
e.V. (\url{www.gauss-centre.eu}) for funding this project by providing
computing time on the GCS Supercomputer SuperMUC-NG at LRZ
(allocations {\tt pn36ge} and {\tt pn36jo}).
This material is based upon work supported by NSF's LIGO Laboratory which
is a major facility fully funded by the National Science Foundation.

\end{acknowledgements}

\appendix

\section{NR-informed relations for ejecta properties}
\label{app:nr-map}

We derive updated NR-informed EOS-insensitive relations for the dynamical ejecta mass $\mej^{\dyn}$,
the dynamical ejecta velocity $v^{\dyn}$, and the disk mass $\mdisk$.
We employ the NR data collected from \citep{Hotokezaka:2012ze,Bauswein:2013yna,
	Dietrich:2015iva,Dietrich:2016hky,Lehner:2016lxy,
	Sekiguchi:2016bjd,
	Radice:2018pdn,
	Vincent:2019kor,Kiuchi:2019lls,Perego:2019adq,Endrizzi:2019trv,
	Bernuzzi:2020txg,Nedora:2020qtd}, as available for each quantity. 
The dataset include 262 NR simulations of non-spinning BNS mergers
spanning the ranges $M\in[2.4,4]~\Mo$,
$q\in [1,2.05]$ and $\lambdat\in [50,3200]$. The ejecta data refer to  
simulations with different degree of physical accuracy
\citep{Nedora:2020qtd}. Some simulations do not include microphysics
and/or neglect neutrino absoption, which are known to be important to
describe the mass ejecta, \eg~\citep{Perego:2017wtu}. Nonetheless, we
employ the entire dataset to provide a conservative constrain within the
Bayesian analysis (see below).

The mass of the dynamical component 
is calibrated using the a factorized empirical analytical form\footnote{Here the notation $\log(.)$ indicates the natural logarithm.},
\begin{equation}
\label{eq:mdyn-fit}
\log\left({\mej^{\dyn}}/{M}\right) = a_0\, G(\nu)\, F(m_1,m_2, \Lambda_1,\Lambda_2)\,,
\end{equation}
that includes effects in the symmetric mass ratio $\nu=m_1 m_2/M^2$,
\begin{equation}
\label{eq:nu-corr}
G(\nu) = 1+g_0 \left(1-4\nu\right)\,,
\end{equation}
and in the BNS parameters,
\begin{equation}
F(m_1,m_2,\Lambda_1,\Lambda_2) = 1 +
b_1 \Lambda_1^\beta + c_1 \left(\frac{m_1}{\Mo}\right)^{\gamma} + 
(1\leftrightarrow2)\,.
\end{equation}
The coefficients $\{a_0,b_{1,2},c_{1,2}\}$
are calibrated on NR data using a differential evolution method.
We find the optimal coefficients to be 
\begin{equation}
\label{eq:mdyn-coeff}
\begin{split}
a_0&= -21 \pm 4\,,\quad  
g_0 = -2\pm 3\,,\\
b_{1} &=0.004\pm 0.001 \,,\quad  
c_{1} = -0.5\pm 0.3\,,  \\
b_{2} &= -0.0025 \pm 0.0009 \,,\quad  
c_{2}  =-0.2\pm 0.5\,,\\  
\beta &= 1/2  \,,\quad  
\gamma  =-1/4\,,\\  
\end{split}
\end{equation}
with a $\chi^2 = 4.91$ and a standard deviation of the residuals equal to $13.6\%$.
For the velocity of the dynamical ejecta,
we employ a similar fitting formula, where the coefficients are 
\begin{equation}
\label{eq:vdyn-coeff}
\begin{split}
a_0&= 0.09\pm 0.06\,,\quad  g_0=-5\pm 2\,,\\
b_{1} &=-0.02 \pm0.01\,,\quad  c_{1} = -0.2\pm 0.9 \,, \\
b_{2} &= 0.01 \pm  0.01\,,\quad  c_{2} = 1.3\pm0.6 \,,  \\
\beta &= 1/2  \,,\quad  \gamma  =1\,,\\  
\end{split}
\end{equation}
with a $\chi^2 = 12.1$ and a standard deviation of the residuals equal to $21.0\%$.

We calibrate the disk mass $\mdisk$ extracted from NR data
using the following fitting formula
\begin{equation}
\label{eq:mdisk-fit}
\log\left(\frac{\mdisk}{M}\right) = a_0\,
\chi(\Lambda_1+\Lambda_2)\, 
F(m_1,m_2,\Lambda_1,\Lambda_2)\,,
\end{equation}
where $\chi(\Lambda_1+\Lambda_2)$
is a correction introduced to for small $\Lambda$'s \citep{Radice:2018pdn},
\begin{equation}
\label{eq:corr-mdisk-fit}
\chi(x)= 1+
\chi_0\,  \left[\frac{1}{\pi}{\rm arctan}\left(\frac{x-\Lambda_0}{\Sigma_0}\right)+\half\right]\,.
\end{equation}
The coefficients $\{a_0,b_{1,2}, c_{1,2},\chi_0,\Lambda_0, \Sigma_0\}$
are calibrated on NR data using a differential evolution method.
We find the optimal coefficients to be
\begin{equation}
\label{eq:mdisk-coeff}
\begin{split}
a_0&=-14 \pm 7 \,,\\
b_{1} &=(5\pm 3 )\times 10^{-6} \,,\quad  c_{1} = -0.5\pm 0.3\,, \\
b_{2} &= (2\pm 1 )\times 10^{-6}\,,\quad  c_{2} = 0.3\pm 0.4\,,  \\
\beta &= 2  \,,\quad  \gamma  =2\,,\\  
\chi_0 &= -1.00\pm 0.01\,, \quad 
\Lambda_0 = 550\pm 30  \,, \quad \Sigma_0=180\pm 30 \,, \\
\end{split}
\end{equation}
with a $\chi^2 = 2.64$ and a standard deviation of the residuals equal to $16.4\%$.
As also shown by previous studies~\citep{Radice:2018pdn},
the total amount of disk mass increases for increasing tidal polarizability
that is consequence of the small compactness of the progenitors, \ie~$m/R \simeq 0.1$.
We assume that a fraction $\xi$ of the total disk mass
contribute to the winds $\mej^{\wind}$, 
\begin{equation}
\label{eq:mdisk-frac}
\mej^{\wind} = \xi \, {\mdisk^{\rm fit}}\,.
\end{equation}
None of the relations proposed here is singular on the calibration
range, and all formulas are stable in the limit $\Lambda_i\to 0$.

As discussed above, these NR relations carry non-negligible
uncertainties, quantifiable at the ${\sim}20\%$ level.
In order to marginalize over these theoretical uncertainties,
we introduce appropriate calibration parameters $\delta_k$ in the PE.
These calibration parameters affect the predictions of the dynamical
ejecta properties as~\citep{Breschi:2021tbm,Breschi:2021xrx}
\begin{align}
\label{eq:mej-recalib}
\log\left({\mej^{\dyn}}/{M}\right)&= (1+\delta_1) \, \log\left({\mej^{\dyn}}/{M}\right)^{\rm fit}\\
\label{eq:vel-recalib}
v^{\dyn} &= (1+\delta_2)\, \left(v^{\dyn} \right)^{\rm fit}\,,
\end{align}
where the superscript ``fit'' denotes a prediction of an NR-informed
relation. The calibration parameters $\delta_{1,2}$ are
assumed to be normally distributed with variance prescribed by the 
residual errors.
The disk mass fraction $\xi$ is taken uniformly distributed within the
range $[0,1]$.

\section*{Data Availability}

{\bajes{}} is an open-source software 
available on \href{https://github.com/matteobreschi/bajes}{\scshape github} 
and on \href{https://pypi.org/project/bajes}{\scshape PyPI}.
For the analyses performed in this work,
we employed the newly released version {${\tt 1.1.0}$}.
\TEOB{} is publicly developed on
\href{https://bitbucket.org/eob_ihes/teobresums/src/master/}{\scshape bitbucket} and available on
\href{https://pypi.org/project/teobresums/}{\scshape PyPI}.
  
\noindent
The GW170817 data are provided by the 
\href{https://www.gw-openscience.org/}{GWOSC}. 
The AT2017gfo data are collected from \cite{Villar:2017wcc}.
The NICER posteriors are taken from the corresponding 
references, \ie~\cite{Miller:2019cac,Miller:2021qha,Riley:2019yda,Riley:2021pdl,Vinciguerra:2023qxq}.
The EOS prior set is available on the NR-GW open data community on 
\href{https://zenodo.org/communities/nrgw-opendata}{\scshape Zenodo}.
The posterior samples presented in
this work will be shared on request to the corresponding author.

\bibliographystyle{aa}

 % if your bibtex file is called example.bib

\end{document}